# In-Plane Mechanical Properties of Ultrathin 2D Hybrid Organic-Inorganic Perovskites

Qing Tu[1,2], Ioannis Spanopoulos[3], Poya Yasaei[1,2], Costas Stoumpos,[3] Mercouri G. Kanatzidis[3], Gajendra S. Shekhawat[2*], Vinayak P. Dravid[1,2*]

1. Department of Materials Science & Engineering, Northwestern University, Evanston, IL 60208, USA

2. Northwestern University Atomic and Nanoscale Characterization Experimental (NU*ANCE*) Center, Northwestern University, Evanston, IL 60208, USA

3. Department of Chemistry, Northwestern University, Evanston, IL 60201, USA

**Abstract**

Two-dimensional (2D) hybrid organic-inorganic perovskites (HOIPs) are new members of the 2D materials family with wide tunability, highly dynamic structural features and excellent physical properties. Ultrathin 2D HOIPs and their heterostructures with other 2D materials have been exploited for study of new physical phenomena and novel device applications. The in-plane mechanical properties of 2D ultrathin HOIPs are critical for understanding the coupling between mechanical and other physical fields and for integrated devices applications. Here we report for the first time the in-plane mechanical properties of ultrathin freestanding 2D lead iodide perovskite membranes and their dependence on the membrane thickness. The in-plane Young's moduli of 2D HOIPs are smaller than that of conventional covalently bonded 2D materials. Both the Young's modulus and breaking strength first decrease and then plateau as the thickness increases from monolayer to 4 layers due to interlayer slippage during deformation. Our results show that ultrathin 2D HOIPs exhibit outstanding breaking strength/Young's Modulus ratio compared to many other widely used engineering materials and polymeric flexible substrates, which renders them suitable for application into flexible electronic devices.

**Introduction**

Semiconducting three dimensional (3D) hybrid organic-inorganic perovskites (HOIPs) formulated as $AMX_3$ (A = small organic cation, M = divalent group 14 element and X = halogen) have drawn a great deal of research attention in the photovoltaic (PV) community due to their high Power conversion efficiency (more than 22% in research labs) and the ease of their solution-based processing method.[1-2] Two-dimensional (2D) layered HOIPs have demonstrated improved stability and promising PV efficiency compared to their 3D counterparts.[3-5] Such 2D HOIPs can be derived from the ordered removal of the M-component from the inorganic framework along certain direction in the parent 3D perovskite structure by intercalating a layer of organic cations[6] (Figure 1). Consequently, 2D HOIPs have structures with alternating organic and inorganic layers, where in the case of the 2D Ruddlesden−Popper family of materials that is studied in this work (see below), the organic part consists of two adjacent organic layers interfacing each other through Van der Waals forces (e.g., dispersion forces) (Figs. 1 and S1).[6-7] He we focus on 2D Ruddlesden−Popper family of materials of the formula $(BA)_2(MA)_{n-1}Pb_nI_{3n+1}$ are BA=butyl ammonium and MA is Methylammonium and n=3. The ease and scalability of solution-based processing methods of high quality 2D HOIP thin films, together with the wide tunability of the materials' properties and structural architecture, by various combinations of organic spacer cations and metal-halide frameworks,[6] have allowed their applications to move beyond PVs, into flexible electronics,[8-9] transistors,[10-11] and thermoelectrics.[12-13]

The extraordinary properties of atomically thin 2D materials (*e.g.*, graphene, transition metal dichalcogenides (TMD), and hexagonal boron nitride (h-BN))[14-15] have triggered remarkable research interest in investigating the physical properties of 2D HOIPs at extreme thickness. Such atomically thin 2D HOIP sheets can be directly synthesized via bottom-up method from solution.[16] On the other hand, the presence of weak Van der Waals interfaces in 2D HOIPs enables them to be mechanically exfoliated into ultrathin layers.[15, 17] Compared to other 2D semiconducting materials, 2D HOIPs remain direct bandgap materials regardless of the crystal thickness and the bandgap can be precisely engineered over a wide range via controlling the structure and thickness.[3] The optical, electrical and transport properties of ultrathin 2D HOIP layers are under intense investigaiton.[16, 18-19] Furthermore, heterostructures by stacking ultrathin 2D HOIPs and other 2D materials have been made for novel applications such as photodetectors, phase change memories and topological insulators.[20] However, the evaluation of the mechanical properties of these ultrathin 2D HOIPs, which is essential for understanding the coupling between mechanical and other physical fields and for designing integrated device applications, remains unexplored so far.

The mechanical properties of 3D HOIP crystals and thin films have been well studied by both experimental and computational methods.[21-25] It has been found that both the static and dynamic mechanical properties of 3D HOIPs are dominated by the metal-halide bond strength while the small organic cations A$^+$ have moderate contribution.[21-24] The measured Young's modulus of HOIPs are found to be significantly lower than other inorganic perovskite oxide materials.[26-27] This "soft" nature has benefited the manufacturing process, where perovskite thin film can be easily formed on surfaces with various morphologies by scalable methods.[22-23] Furthermore, this "soft" feature of HOIPs leads to low resistance to structural changes, including phase transition, ion migration and self-healing.[6, 8, 28-29] Recently, we reported the first study of out-of-plane mechanical properties of 2D HOIPs as a function of structural parameters (*i.e.*, the spacer molecule R and number of inorganic layers n sandwiched by two organic layers in the crystal structure).[30] We found that the presence of the flexible organic layer and its Van der Waals interfaces further softens the material, while a variety of organic cations can be used to precisely engineer the out-of-plane mechanical properties, by tuning the stiffness of the organic layer and the interfacial interaction strength.

While the out-of-plane mechanical property of monolayer 2D materials highly depends on the materials they interface with,[31-33] their in-plane mechanical property is intrinsic to the in-plane bond strength and atomic structures.[34] Recently, it was revealed that the sliding energy between the Van der Waals bonded layers dictated by the interlayer interaction strength has a significant impact on the in-plane mechanical property of few-layer 2D materials.[35-37] For example, 8-layer graphene has a Young's modulus about 0.94 TPa, less than that of monolayer graphene (~ 1.03 TPa), while 9-layer h-BN has a similar Young's modulus value compared to monolayer h-BN (0.86 TPa versus 0.87 TPa, respectively).[35]

The most widely used experimental method for measuring the in-plane mechanical properties of low-dimensional nanomaterials is atomic force microscopy (AFM) indentation of the suspended samples,[34] where the load applied by AFM tip causes mainly in-plane stretching of the suspended materials. This approach can be dated back to the study of nanowires and carbon nanotubes,[38-39] and its application to 2D materials starts from the seminal work on mechanical exfoliated graphene.[40] Since then, it has been utilized to investigate the mechanical behavior of various chemical vapor deposited (CVD) or mechanically exfoliated 2D materials and heterostructures that are covalently bonded in plane.[35, 37, 41-43]

Here we report the study of the in-plane mechanical properties of ultrathin 2D lead iodide HOIPs, which has ionic nature in-plane,[6, 20] and their dependence on the thickness of the materials by the well-established AFM-based indentation technique. We find both Young's modulus and breaking strength first decrease and then plateau as the membrane thickness increases from monolayer to 4 layers. We compare our results to other 2D materials with covalent bonds in-plane and provide insights into tuning the mechanical properties of few-layer 2D HOIPs. Our results reveal that ultrathin 2D HOIPs can sustain large deformation, thus being very promising for flexible electronic applications.

**Results and Discussions**

**Preparation and Characterization**. 2D Ruddlesden-Popper HOIP crystals with chemical formula of $(CH_3(CH_2)_3NH_3)_2(CH_3-NH_3)_2Pb_3I_{10}$ (C4n3) were synthesized following the protocols we developed before[3, 30] (See Materials and Methods for more details). For simplicity purpose, we use the notation C4n3 to represent this 2D HOIPs. The number after C indicates the number of carbon atoms in the organic spacer molecule chain, and the number right after n indicates the number of the inorganic $[PbI_6]^{4-}$ layers that are separated by the organic layers (See Supporting Information (SI) – Figure S1 for a schematic of the crystal structure). Silicon wafers with 300 nm $SiO_2$ were patterned with holes (~ 500 nm in depth, 1 μm or 1.5 μm in diameter) by lithography (See Materials and Methods for details). Due to the weak Van der Waals interaction between the two interfacing organic layers (Fig. 1B), C4n3 can be mechanically exfoliated by scotch tapes and then transferred to the hole-patterned silicon wafer, similar to other 2D materials like graphene[40] and $MoS_2$.[41] Thin C4n3 flakes were first identified by optical microscopy (Fig. 2A) and then imaged by tapping mode AFM to measure the thickness (Fig. 2B). The monolayer, bilayer, trilayer and tetralyer C4n3 thickness were measured to be 2.7 ± 0.2 nm, 5.3 ± 0.2 nm, 7.7 ± 0.3 nm, 10.6 ± 0.3 nm (Figs. 2C and S2), respectively, which match well to the reported values and the crystal structures.[3] Intact mono- and few-layer C4n3 membranes suspended on patterned holes could be located by tapping mode AFM without being damaged (Fig. 2B). Similar to other 2D materials,[35, 40, 43] ultrathin C4n3 membranes were well adhered to the edges of the holes due to Van der Waals forces, showing a slightly lower height at the hole edges (Fig. 2B).

**Mechanical Test by AFM Nanoindentation**. Mechanical properties of ultrathin C4n3 were measured by nanoindentation using AFM with a standard AFM probe in ambient environment (See Materials and Methods for details). The samples used were freshly prepared and measured within 24 hours. The spring constant and deflection sensitivity of the cantilevers used in the experiments were calibrated by standard thermal method (See Materials and Methods for more details).[44] For each hole, an AFM topographic image of the suspended C4n3 membrane was obtained by tapping mode and used to position the tip to the center of the membrane for nanoindentation measurements.[35, 40]

For nanoindentation, the force ($F$) and the distance of piezo moving in vertical directions ($Z_{piezo}$) were measured while the probe was moving at the rate of 100 nm/s. Membranes without obvious sliding features and hysteresis in the loading and unloading curves were selected for mechanical properties analysis.[35, 40, 45] The measured $Z_{piezo}$ were converted into membrane displacement ($\delta$) following the methods established in graphene and MoS$_2$ studies:[35, 40, 43]

$$\delta = Z_{piezo} - \delta_{tip}, \tag{1}$$

where $\delta_{tip}$ is the deflection of the AFM tip. Figure 3 shows representative $F - \delta$ curves of C4n3 membrane with different thickness. Here we model the deformation of suspended ultrathin 2D C4n3 membrane as an isotropic thin film. The mechanical deformation of such free-standing thin film can be divided into 2 regimes. Under relatively small load, the stiffness arising from pre-tension due to Van der Waals interaction of the film to the substrate is dominant and the force-displacement should be linear. At large load, the bending stiffness of the suspended thin film starts to take over and force-displacement relationship becomes cubic.[40, 46-47] Consequently, the total $F - \delta$ relationship of the 2D C4n3 membrane should include both linear and cubic part and can be modeled as:[40, 46-47]

$$F = \sigma_0^{2D}(\pi a)\left(\frac{\delta}{a}\right) + E^{2D}(q^3 a)\left(\frac{\delta}{a}\right)^3, \tag{2}$$

where $a$ is the radius of the hole measured by AFM topographic images; $q = 1/(1.05 - 0.15\upsilon - 0.16\upsilon^2) = 1.0095$ is a dimensionless number and $\upsilon = 0.3$ is the Poisson's ratio of lead iodide HOIPs.[22-23] $\sigma_0^{2D}$ is the pretension in the ultrathin 2D C4n3 membranes; $E^{2D}$ is the 2D effective Young's modulus of the membranes, which can be converted into the conventional 3D bulk Young's modulus $E$ by dividing it by the thickness of the membrane. For m layers of 2D C4n3, the effective thickness is taken as m × 2.625 nm (*i.e.*, the distance between 2 neighboring Van der Waals interfaces (see Figs. 1B and S1).[3] The red solid lines in Figure 3 are the least-squares curve fit of the experimental data based on Equation (2), taking $\sigma_0^{2D}$ and $E^{2D}$ as free parameters. The great agreement between the fitted curves and experimental data throughout the tested range confirms the appropriateness of the model. Note that we had two hole sizes but we did not see differences in mechanical properties for C4n3 with the same thickness on different hole sizes. Over 10 membrane samples were measured for each thickness to obtain the average values of mechanical properties.

**Elastic Modulus and Breaking Strength.** The extracted pre-tension $\sigma_0^{2D}$ in the ultrathin 2D C4n3 falls in the range of 0.01 to 0.1 N/m. The pretension of suspended 2D materials depends on both their intrinsic mechanical properties and transferring process. $\sigma_0^{2D}$ in these ultrathin 2D C4n3 membranes are similar to those in dry-transferred CVD MoS$_2$[43] and wet-transferred CVD graphene,[48] but smaller than the pretension in mechanical exfoliated graphene[40] and MoS$_2$.[41] Figure 3 summarizes the obtained Young's moduli of C4n3 with different number of layers. Monolayer has an $E^{2D}$ = 29.4 ± 3.6 N/m, which about 1 order of magnitude smaller than those from 2D materials bonded covalently in-plane such as graphene,[35, 40] MoS$_2$[37, 41, 43] and h-BN,[35] which agrees well with the fact that covalent bonds are usually much stronger than ionic bonds.[20] The corresponding bulk Young's modulus $E$ in the in-plane direction is 11.2 ± 1.4 GPa. This value falls in the range of experimentally measured Young's modulus of 3D $CH_3NH_3PbI_3$ single crystal along <100> direction by nanoindentation,[23-24] where the whole crystal consists of $PbI_6$ octahedra (Fig. 1A). This suggests the obtained Young's modulus represents the intrinsic Young's modulus of 2D lead iodide perovskites. On the other hand, $E$ of the monolayer C4n3 is about 2.5 times of the Young's modulus of C4n3 crystals in the out-of-plane directions where the weak Van der Waals interaction and soft organic spacing molecules -$C_4H_9$ significantly lowers the stiffness.[30] However, monolayer C4n3 is much softer than graphene, $MoS_2$ and h-BN, whose moduli are about 20 to 100 times higher.[35, 37, 40-41, 43] Compared to the planar structure connected through covalent bonds in-plane in these 2D materials, the $[PbI_6]^{4-}$ layers in C4n3 have much higher deformability. Such crystal structure induced softness in the mechanical properties have been found in black phosphorous (BP), where the "atomic puckers" along the zig-zag direction can be relatively easily deformed by tensional stresses along the armchair direction.[36, 42] The softness of 2D HOIPs can lower the energy barrier of structural changes like phase transition under pressure,[6, 49] which can lead to novel applications such as phase change memories and topological insulators.[20, 50]

If the bulk elastic modulus $E$ of the 2D materials is independent of the 2D membrane's thickness, $E^{2D}$ should increase linearly as a function of the thickness, as what has been reported for h-BN.[35] However, the moduli of thicker C4n3 show a clear thickness dependence and significantly deviate from the linear projection (the red dashed lines in Figs. 4A and B). Bilayer C4n3 has a $E^{2D}$ = 37.1 ± 4.9 N/m, corresponding to $E$ = 7.1 ± 0.9 GPa, which decreases by 37% compared to that of the Young's modulus of monolayer C4n3. Such decrease in Young's modulus from monolayer to bilayer have also been reported in graphene (~ 6%),[35] BP (~ 15%)[36] and MoS$_2$ (~ 26%),[43] but not in h-BN,[35] which are attributed to interlayer sliding arising from the weak interlayer interaction in graphene, BP and MoS$_2$ compared to that in h-BN.[35-36, 43] Therefore, the observed thickness dependence of ultrathin C4n3 Young's modulus also suggests the presence of the inter-layer sliding at the Van der Waals interfaces during AFM induced deformation due to the weak lateral interlayer interaction in this type of HOIPs. Such interlayer sliding can happen as a result of shift in interdigitation in the crystal structure and/or shear deformation of the flexible alkyl carbon -C$_4$H$_9$ chains in the spacer molecules under mechanical loading.

As the monolayer thickness further increases to 3 layers and 4 layers, $E$ of C4n3 plateaus around 5.7 GPa. The in-plane Young's modulus for these thick few-layer C4n3 is slightly higher than that in the out-of-plane direction of the same crystals.[30] The thickness independence of mechanical properties at large thickness (> 3 layers) has also been found in other 2D materials, showing the saturation of the sliding effects on the mechanical properties.[35-36, 42] We further measured the modulus of ultrathin HOIPs with a general formula (CH$_3$-(CH$_2$)$_3$-NH$_3$)$_2$PbI$_4$ (abbreviated as C4n1 below), which has similar structure to C4n3 but contains only 1 inorganic [PbI$_6$]$^{4-}$ anionic layer sandwiched by two organic CH$_3$-(CH$_2$)$_3$-NH$_3^+$ layers (See Fig. S1). Due to the small mechanical strength of C4n1, it is very challenging to obtain intact membrane with thickness less than 8 nm suspended over the holes. The thinnest freestanding C4n1 membrane we can get is about 8.2 ± 0.2 nm, which is about 6 repeating monolayers of C4n1.[3, 16] As shown in Fig. 4B, 6-layer C4n1 has an in-plane Young's modulus about 5.7 ± 0.9 GPa, almost equal to the plateau Young's modulus of few-layer C4n3 HOIPs. 6-layer C4n1 contains a total of 6 inorganic $[PbI_6]^{4-}$ octahedral layers (similar to that of bilayer C4n3) but also 5 organic Van der Waals interfaces (more than that of 4-layer C4n3). This suggests that the sliding effect between layers indeed saturates at large thickness and the in-plane elastic modulus plateaus at about 5.7GPa for thick C4n3 HOIPs, which is also significantly lower than the literature reported values of 3D $CH_3NH_3PbI_3$ single crystal along <100> direction by nanoindentation.[22-23, 51] For graphene, $E$ reaches the plateau at about 4 layers and the modulus value of 4-layer graphene is about 92% of that of monolayer graphene.[35] For BP, $E$ plateaus at about 3 layers and 3-layer BP has moduli values about 96% and 84% of those of monolayer BP along the zig-zag and armchair directions, respectively.[36] Compared to graphene and BP, the presence of Van der Waals interfaces causes much more reduction in Young's modulus in C4n3 HOIP. This indicates a much smaller energy barrier against interlayer sliding in 2D C4n3 HOIP membranes, which suggests that C4n3 can potentially be used as a good solid lubricant[52-53] Furthermore, our study also suggests that the in-plane elastic modulus of 2D HOIPs can be further tuned by selecting organic cations to form stronger or weaker interlayer Van der Waals interactions.

We further increase the forces to break the suspended C4n3 membrane. Our experiments are in the limit of small indenter characterized by $r_{\text{tip}}/a \ll 1$ ($r_{\text{tip}}/a \approx 0.04$ or 0.025 in our case, for small or large holes, respectively) and large loads with respect to pretension, as determined by a factor $\kappa \ll 1$ defined as[54]

$$\kappa = \left(\frac{\sigma_0}{E^{2D}}\right)^{1/3} \left(\frac{2r_{\text{tip}}}{a}\right)^{2/3} \tag{3}$$

In our case, $\kappa < 0.03$. Thus we can calculate the breaking strength of C4n3 using the expression for maximum stress $\sigma_m^{2D}$ in a clamped, linear elastic, circular membrane under a spherical indenter:[54]

$$\sigma_m^{2D} = \sqrt{\frac{F_m E^{2D}}{4\pi r_{\text{tip}}}}, \tag{4}$$

where $F_m$ is the maximum force at which the membrane breaks. Equation (4) has been used to evaluate the breaking strength other 2D materials like MoS$_2$[43] and BP.[42] The average maximum stress for monolayer C4n3 membrane can sustain is 1.9 N/m with a standard deviation of 0.2 N/m (Fig. 5A), which corresponds to a 3D breaking strength of 0.7 ± 0.08 GPa (Fig. 5B). On average, this corresponds to about 5 to 8% of the Young's modulus of monolayer C4n3 HOIP. These values are very close to the theoretical upper limit of a material's breaking strength, which suggests that the solution grown C4n3 HOIP crystals have quite low defect densities. The breaking strength/Young's modulus of C4n3 is lower than that of graphene,[40] but comparable to those of MoS$_2$,[43] WS$_2$ nanotubes[55] and carbon nanotubes,[56] and larger than those of many widely used engineering materials including Kevlar 49,[57] polydimethylsiloxane (PDMS),[58] and stainless steel,[59] which highlights the outstanding mechanical performance of this material.

Finally, we investigated the breaking strength as a function of thickness. Similar to the elastic modulus, we found that the 2D breaking strength $\sigma_m^{2D}$ of C4n3 deviates significantly from the projection obtained by multiplying that of monolayer C4n3 by the layer numbers (red dashed line in Fig. 5A). The converted 3D bulk breaking strength $\sigma_m$ decreases and then plateaus (Fig. 5B) as the thickness of the ultrathin C4n3 membrane increases. This trend in breaking strength is very similar to that found in few-layer graphene,[35] MoS$_2$/graphene and MoS$_2$/WS$_2$ heterostructures,[37] which further verified the occurrence of interlayer slippage during mechanical deformation. The breaking strength of C4n3 HOIPs eventually converges to about 0.36 GPa. Considering the plateau value of Young's modulus, this suggests that few-layer C4n3 can sustain strain larger than 6%. In comparison, thin polymer films that are commonly used as substrates for flexible electronics, such as polyimide, poly(ethylene terephethalate), and aromatic polyamide, break or yield (*i.e.*, plastic deformation) at a strain about 6~7%,[60-61] which is close to the values of mono- and few-layer C4n3 HOIPs. This suggests that 2D C4n3 HOIPs can be readily integrated with these substrates for applications in flexible electronics. Clearly, the strength of the Van der Waals interfacial interaction determines the extent of interlayer slippage in few-layer 2D materials, which has an impact on the in-plane mechanical properties of few-layer 2D materials. However, in most 2D materials like graphene and MoS$_2$, such interfacial interaction is completely determined by the nature of the 2D layers and is not tunable. However, unlike other 2D materials, the huge variety of organic spacer molecules that can be incorporated into 2D HOIPs allows for much more latitude in engineering the strength of the Van der Waals interfacial interaction and thus tailoring the mechanical properties of 2D HOIPs to fit the need of different applications.

**Summary & Conclusions**

In conclusion, we measured the in-plane mechanical properties of suspended ultrathin C4n3 HOIP membranes and their dependence on the membranes' thickness. We find that monolayer C4n3 has an in-plane Young's modulus larger than that in the out-of-plane direction. These ionically bonded 2D materials are much softer than other covalently bonded 2D materials. As the thickness increases, both the Young's modulus and breaking strength first decrease and then plateau due to interlayer slippage during deformation. The latter arises from the weak interfacial interactions between the organic layers. Our results suggest that this family of 2D compounds could be potential candidates for investigations as solid lubricants. The hybrid nature and wide tunability of 2D HOIPs should allow the engineering of the interfacial interaction strength to tailor their in-plane mechanical properties. The breaking strength/Young's modulus ratio of ultrathin 2D C4n3 HOIPs is very high, which suggests low defect density and high crystal quality in these solution grown crystals. These unique mechanical properties of ultrathin 2D C4n3 HOIPs are outstanding compared to those of the widely used engineering materials (e.g., stainless steel and Kevlar 49) and flexible polymeric substrates (e.g., polyimide), thus satisfying the mechanical requirements for commercially viable flexible electronic applications.

**Materials and Methods**

*Materials.* We synthesized the 2D C4n3 and C4n1 HOIP crystals following the protocol we developed before (see SI-Section XX for details).[3] The synthesized 2D HOIP samples were characterized by powder X-ray Diffraction measurements (PXRD) and the resulting patterns were compared with the calculated ones from the solved single crystal structures (See Fig. S3) to verify the phase purity. Extensive analysis of the single crystal structures of the 2D HOIPs can be found elsewhere.[3]

*Suspended Ultrathin C4n3 Membrane Preparation.* To fabricate the hole arrays, 50 nm chromium (Cr) was first deposited on a silicon wafer (having 300 nm thermally-grown SiO2) with e-beam evaporation. A stepper was used to pattern the hole arrays on the substrate through standard lithography recipes. Cr was etched in Transene Cr Mask Etchant 9030 for 75 seconds and rinsed in DI water. Photoresist was then removed in N-Methyl-2-pyrrolidone (NMP) for 2 hours. Next, using the patterned Cr hard mask, $SiO_2$ and silicon were etched in an inductively coupled plasma (ICP) Fluorine etch system to a total depth of ~0.5 µm. The chromium hard mask was then fully removed by soaking in Cr etchant for 5 minutes followed by solvent cleaning processes. Before transferring the 2D HOIP layers, the patterned silicon wafer was further cleaned by freshly prepared Piranha solution (volume ratio 3:1 for 98% $H_2SO_4$ and 35% $H_2O_2$), then rinsed thoroughly by DI water and blown dry by $N_2$ flow. Ultrathin C4n3 layers were mechanically exfoliated and transferred to the patterned silicon wafer by scotch tapes. The fabricated samples were characterized and measured on the same date.

***AFM measurements.*** All AFM measurements were conducted in ambient environment with an Icon AFM (Bruker, CA). Prior to the measurements, the deflection sensitivity of the AFM cantilever (SCM-PIT-V2, Bruker, CA) was calibrated by engaging the cantilever on a clean silicon surface. The spring constant $k_c$ of the cantilever was calibrated by fitting the first free resonant peak to equations of a simple harmonic oscillator[62] to measure the power spectral density of the thermal noise fluctuations in air.[44] The tip radius is measured by scanning electron microscopy (SEM). The calibrated spring constants and tip radius of the cantilevers used in the current study falls in the range of 3.2 to 3.5 N/m, and 17 to 29 nm, respectively. These values are very close to the nominal values provided by the manufacturer.

**Supporting Information**

Crystal structure schematics of C4n3 and C4n1, Details about synthesis and XRD characterization, and AFM images of bilayer, trilayer, 4-layer C4n3 membranes and 6-layer C4n1 membranes.

**Author Information**

*Corresponding authors:

Dr. Gajendra S. Shekhawat: g-shekhawat@northwestern.edu

Prof. Vinayak P. Dravid: v-dravid@northwestern.edu

**Acknowledgement**

The work made use of the SPID and EPIC facilities of Northwestern University's NUANCE center, which has received support from the Soft and Hybrid Nanotechnology Experimental (SHyNE) resource (NSF ECCS-1542205); the MRSEC program (NSF DMR-1720139) at the Materials Research Center; the International Institute for Nanotechnology (IIN); the Keck Foundation, and the State of Illinois, through the IIN. This work was supported by the National Science Foundation IDBR Grant Award Number 1256188, and partially supported by Air Force Research Laboratory grant FA8650-15-2-5518. M.G.K. acknowledge the support under ONR Grant N00014-17-1-2231.

**Figures**

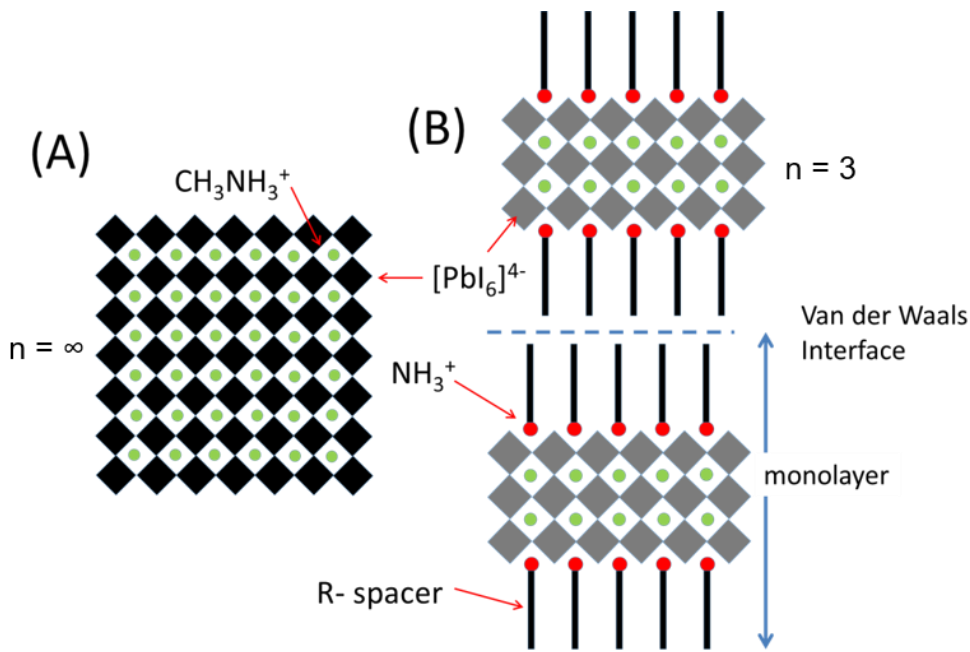

**Figure 1.** Schematics illustrating the dimensional reduction of 3D AMX$_3$ HOIPs to 2D (R-NH$_3$)$_2$(A)$_{n-1}$(M)$_n$(X)$_{3n+1}$ HOIPs. A) represents the 3D HOIP framework where by ordered removal of the inorganic [PbI$_6$]$^{4-}$ layer and its replacement by the organic spacer molecules (R-NH$_3^+$) to get the 2D structure (B). In this specific example, A = CH$_3$NH$_3^+$, M = Pb, and X = I. and the 2D structure consists of 3 inorganic perovskite layers and one organic layer below and above them. In the current study, R = CH$_3$-(CH$_2$)$_3$.

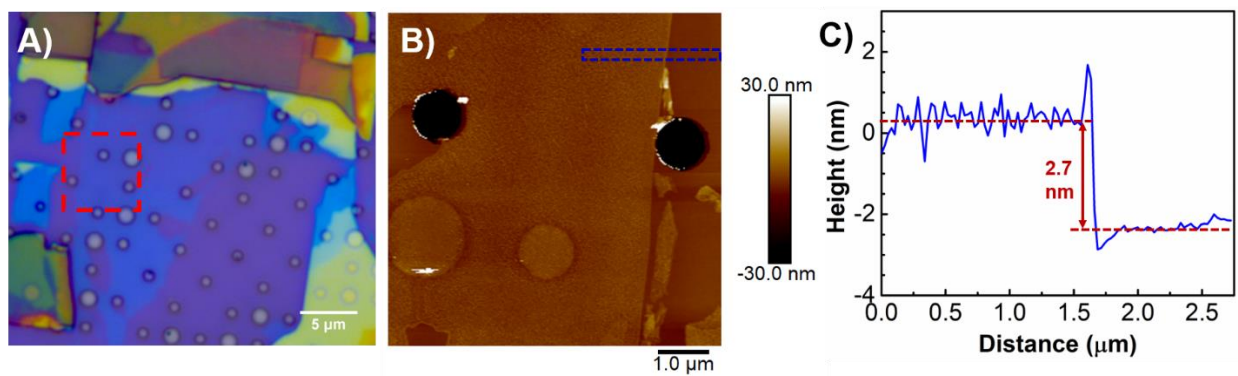

**Figure 2.** Characterization of a monolayer C4n3 flake: A) optical image of a monolayer C4n3 flake on a SiO$_2$/Si substrate with holes (~ 1 μm and 1.5 μm in diameter); B) AFM topographic image corresponding to the red box in (A), and C) is height profile corresponding to the blue box in (B).

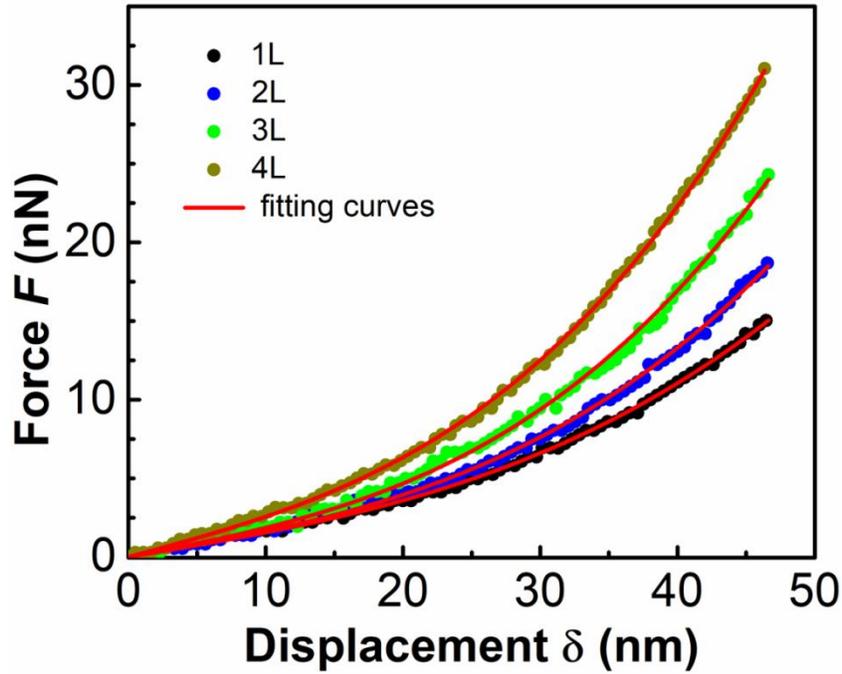

**Figure 3.** Force – displacement (F-δ) curves of C4n3 membranes with different thickness with fitting by Equation (2).

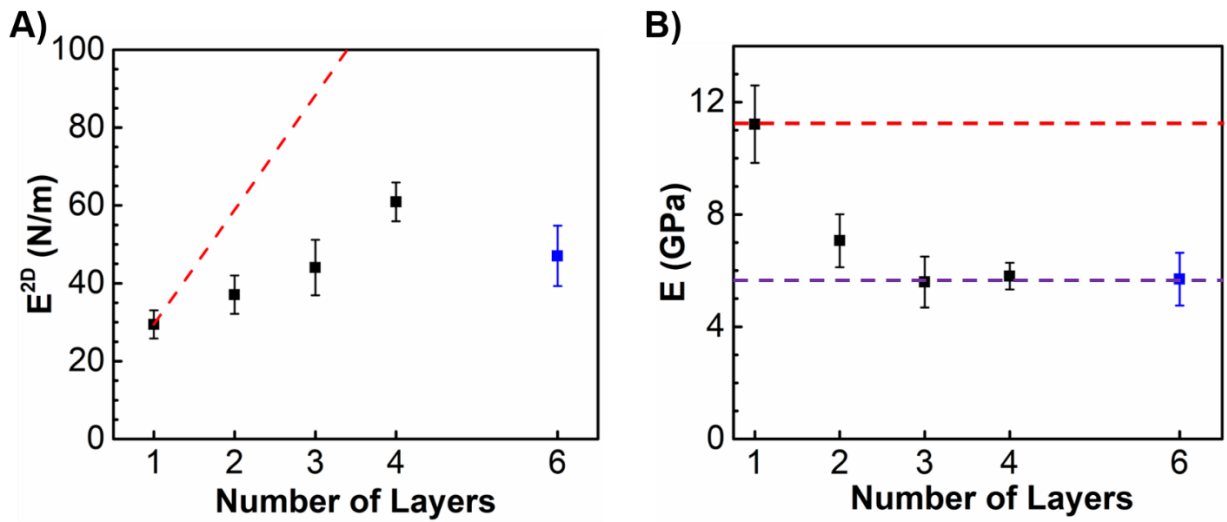

**Figure 4.** Elastic properties of ultrathin C4n3 membranes: A) and B) 2D and 3D bulk Young's modulus of C4n3 as a function of thickness, respectively. The red dashed line represents expected trend without interlayer sliding. Blue dot is value for 6-layer C4n1 perovskite. Purple dashed line indicates the plateau value.

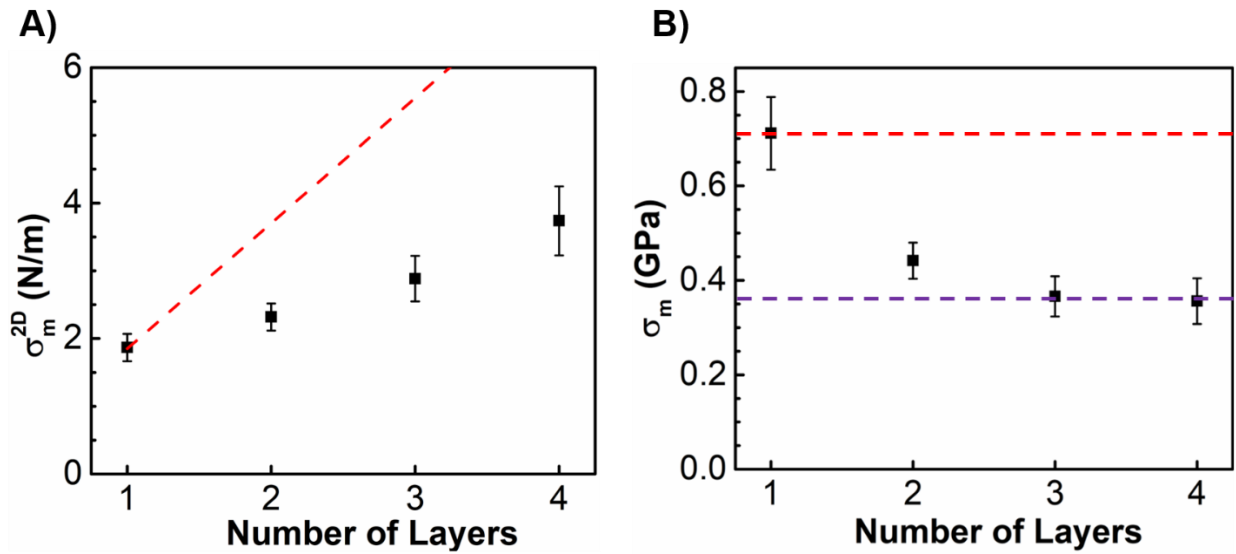

**Figure 5.** Mechanical strengths of ultrathin C4n3 membranes: A) and B) 2D and 3D breaking strength of C4N3 with different thicknesses, respectively. The red dashed line represents expected trend without interlayer sliding. Purple dashed line indicates the plateau value.

**Table of Content**

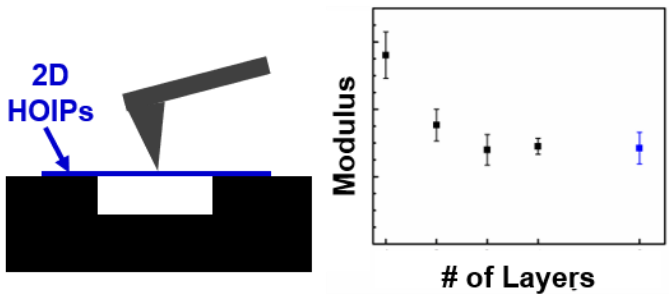

23